\documentclass[11pt,showpacs,preprintnumbers,amsmath,amssymb]{revtex4}
\usepackage{graphicx}% Include figure files
\usepackage{dcolumn}% Align table columns on decimal point
\usepackage{bm}% bold math
\usepackage{mathrsfs}
\usepackage{amssymb}
\usepackage{amsmath}
\usepackage{amsfonts}
\usepackage{amsfonts,enumerate}
\usepackage{pifont}
\usepackage{enumerate}
\usepackage{graphics}
\usepackage{verbatim}
\usepackage{amsthm}

\begin{document}

%\preprint{APS/123-QED}

\title{Hawking radiation as tunneling is non-Markovian: An information-theoretic solution
to the paradox of black hole information loss}% Force line breaks with \\

\author{Zeqian Chen}
\email{zqchen@wipm.ac.cn}
\affiliation{%
State Key Laboratory of Resonances and Atomic and Molecular Physics, Wuhan Institute of Physics and Mathematics,
Chinese Academy of Sciences, 30 West District, Xiao-Hong-Shan, Wuhan 430071, China}%

\date{\today}% It is always \today, today,
             %  but any date may be explicitly specified

%\begin{abstract}
%In this paper
%\end{abstract}

% PACS, the Physics and Astronomy
                             % Classification Scheme.
%\keywords{Hawking radiation as tunneling, non-Markovian, non-classical correlation, entropy}%Use showkeys class option if keyword
                              %display desired
\maketitle

In an ontological description of a physical system \cite{Spekkens2005}, the state evolution of the system is called {\it Markovian} if the probability distributions associated with the states of the system at all time in the future depend only on the probability distribution associated with the present state but not on previous ones \cite{Montina2008}; ``{\it non-Markovian}" is the opposite. Given a system and a set of initial conditions, classical mechanics allows us to calculate the future evolution to arbitrary precision. Any uncertainty we might have at a given time is caused by a lack of knowledge about the configuration. This indicates that the state (deterministic or stochastic) evolution in classical mechanics has to be Markovian, that is, a probability distribution (or knowledge) for the actual state of a classical system at any time in the future actually only depends on the knowledge for the present state but not on probability distributions of the past ones. As is well-known, Hawking's 1974 announcement \cite{Hawking1974} that black holes evaporate thermally has led to a much-debated paradox that the consequent loss of information violates the principle of unitarity for quantum mechanics. This paradox has attracted a number of proposals, but no consensus on a solution \cite{HS2010}. Here, we show that Hawking radiation as tunneling \cite{KW1995} is non-Markovian in terms of the tunneling probability uncovered by Parikh and Wilczek \cite{PW2000}. This implies that there are non-classical correlations among Hawking radiations. By the Bayesian rule in information theory \cite{CT1991}, we compute the amount of information encoded in those correlations and find that it is equal to the maximum information content of a black hole measured by the Bekenstein-Hawking entropy \cite{BH1973}. By noting that the Bekenstein-Hawking entropy is a measure of the hole's information capacity \cite{IY2010}, we arrive at the conclusion that there is no information loss in Hawking radiation, and thereby provide an information-theoretic solution to the paradox of black hole information loss.

Recently, Zhang {\it et al} \cite{ZCYZ2009} have shown that there exist statistical correlations between quanta of Hawking radiation as tunneling based on the principle of energy conservation involving the accepted emission probability spectrum from a black hole, as conjectured by Parikh and Wilczek \cite{PW2000}. They concluded that up to $\mathrm{exp} S_{\mathrm{BH}}$ bits of information can carried off in the correlation, which is the maximum information context of a black hole measured by the Bekenstein-Hawking entropy $S_{\mathrm{BH}}$ in the sense that $\mathrm{exp} S_{\mathrm{BH}}$ is the maximum number of bits that can be accommodated in a black hole formed by an astrophysical collapse. As pointed out in \cite{IY2010}, this goes a considerable way toward resolving this long-standing ``information loss paradox."

In this article, we offer a contribution to the solution of the problem by approaching the different options on the table with an information-theoretic argument. For simplicity, we take the convenient units of $k = \hbar=c = G =1.$ Suppose to consider Hawking radiation as tunneling for the Schwarchild black hole. The tunneling probability for an emission at an energy $E$ is found to be \cite{PW2000}
\begin{equation}\label{eq:EmissionProb}
P(E) = \exp \Big [ -8 \pi E \Big ( M - \frac{E}{2} \Big ) \Big ] \equiv \exp ( \triangle S),
\end{equation}
where the second equal sign expresses this result in terms of the change of the Bekenstein-Hawking entropy for the Schwarchild black hole $S_{\mathrm{BH}} = A /4 = 4 \pi M^2$ with $A= 4\pi (2 M)^2$ being the surface area of a Schwarzchild black hole with mass $M$ and radius $2 M.$ Let us consider two emissions arising simultaneously with energies $E_1$ and $E_2$ respectively. According to an information-theoretic (Bayesian) view, in which probabilities are primarily states of knowledge or evidence, the probability $P(E_1, E_2)$ of such two emissions is equal to the probability of an emission at an energy $E_1 + E_2,$ that is,
\begin{equation}\label{eq:TwoEmissionProb}
P(E_1, E_2) = P(E_1 + E_2) = \exp \Big [ -8 \pi (E_1 + E_2) \Big ( M - \frac{1}{2} (E_1 + E_2) \Big ) \Big ],
\end{equation}
because the black hole emits the same energy in both cases and so the information we can obtain are the same. This inference clearly can apply to the case of many emissions arising simultaneously and the corresponding probability formula holds true.

Now consider the first three sequential emissions with energies $E_1, E_2,$ and $E_3$ in a sequence of emissions. If the state evolution of the total system composed of a black hole and its radiations as tunneling were Markovian, then the conditional probabilities $P(E_3 | E_2, E_1)$ and $P(E_3 | E_2)$ should satisfy
\begin{equation}\label{eq:Markov}
P(E_3 | E_2, E_1) = P(E_3 | E_2).
\end{equation}
However, by the Bayesian rule we find that \cite{ConditionalProb}
\begin{equation*}
\ln P(E_3 | E_2, E_1) - \ln  P(E_3 | E_2) = 8 \pi E_1 E_3 \neq 0.
\end{equation*}
Thus Hawking radiation as tunneling has to be non-Markovian which shows that there are non-classical correlations among these radiations \cite{ClassicalCorre}.

Next we compute the information content hidden in those correlations of sequentially tunneled particles, that is, the entropy of all tunneled particles. Let us consider a radiation process of sequential emissions with energies $E_1, E_2, \ldots, E_n$ so that $M = \sum^n_{i=1} E_i.$ According to the Chain rule for entropy in information theory \cite{CT1991}, the total entropy $H(E_1, E_2, \ldots, E_n)$ of this radiation process which eventually exhausts the black hole is
\begin{equation}\label{eq:ChainEntropy}
H(E_1, E_2, \ldots, E_n) = \sum^n_{i =1} H (E_i | E_{i-1}, \ldots, E_1)
\end{equation}
where $H (E_i | E_{i-1}, \ldots, E_1) = - \ln P (E_i | E_{i-1}, \ldots, E_1)$ with $H (E_1 | E_1) = - \ln P(E_1).$ Again, by the Bayesian rule we find that \cite{ConditionalEntropy}
\begin{equation}\label{eq:TotalEntropy}
H(E_1, E_2, \ldots, E_n) = 4 \pi M^2,
\end{equation}
which is exactly the same as the Bekenstein-Hawking entropy of the Schwarzchild black hole with mass $M$ and radius $2 M.$

In conclusion, we have shown that the information flow associated with the state evolution of the total system composed of a black hole and its radiations as tunneling is non-Markovian and thus there are non-classical correlations among these radiations. We further find that the entropy of the system is conserved if information hidden in those correlations of sequentially tunneled particles is included. Our analysis relies on an information-theoretic argument within which no conservation law (such as the principle of energy conversation) is involved. The crucial feature of our approach is that it is operational, in the sense that we only refer to directly information-theoretic objects, but do not assume anything about the underlying evolution of black holes (such as the energy change of the black hole after an emission). This thereby provides a complete solution to the paradox of black hole information loss from an information-theoretic viewpoint \cite{Extension}. Finally, we remark that there is a close relationship between non-Markovian and unitary evolutions \cite{non-MarkUnitary}. However, it is not our purpose to enter into this debate here.

The author is grateful to Qing-yu Cai for helpful discussions on this topic. This work was supported in part by the NSFC under Grant No. 11171338 and National Basic Research Program of China under Grant No. 2012CB922102.

%\newpage %Just because of unusual number of tables stacked at end
\bibliography{apssamp}% Produces the bibliography via BibTeX.

\begin{thebibliography}{**}
\bibitem{ConditionalProb} Indeed, by the Bayesian formula we have the conditional probability $P(E_3|E_2) = P (E_2, E_3) / P(E_2) = P(E_2 + E_3)/ P(E_2)$ where the second equality is obtained by \eqref{eq:TwoEmissionProb}. Similarly, we have $P(E_3 | E_2, E_1) = P(E_1 +E_2 + E_3)/P(E_1 +E_2).$
\end{thebibliography}

\end{document}